\documentclass[%
reprint,
showpacs,preprintnumbers,
amsmath,amssymb,
aps,
]{revtex4-1}
\pdfoutput=1
\usepackage{graphicx}
\usepackage{dcolumn}
\usepackage{bm}
\usepackage{siunitx}
\usepackage{lineno}
\usepackage{array}
\usepackage{multirow}

\newcommand{\mydecayheader}{\boldmath $e^+e^- \rightarrow \eta_c\eta\pi^+\pi^-$ }
\newcommand{\mydecay}{$e^+e^- \rightarrow \eta_c\eta\pi^+\pi^-$ }
\newcommand{\uplimone}{6.2}
\newcommand{\uplimtwo}{10.8}
\newcommand{\uplimthree}{27.6}
\newcommand{\uplimfour}{22.6 }
\newcommand{\uplimfive}{23.7 }
\newcommand{\Ks}{$K^0_S$}
\newcommand{\dbs}{\bar{D}^*}
\newcommand{\GeV}[1]{#1~\mathrm{GeV}}

\begin{document}
	
	\title{}
	\author{}
	\begin{titlepage}
		\begin{center}
			\textbf{Search for the reaction channel \mydecayheader at center-of-mass energies from $4.23$ to $4.60$ GeV} 
		\end{center}
	    \begin{small}
\begin{center}
M.~Ablikim$^{1}$, M.~N.~Achasov$^{10,d}$, P.~Adlarson$^{59}$, S. ~Ahmed$^{15}$, M.~Albrecht$^{4}$, M.~Alekseev$^{58A,58C}$, A.~Amoroso$^{58A,58C}$, F.~F.~An$^{1}$, Q.~An$^{55,43}$, Y.~Bai$^{42}$
% ...
\\
\vspace{0.2cm}
(BESIII Collaboration)\\
\vspace{0.2cm} {\it
$^{1}$ Institute of High Energy Physics, Beijing 100049, People's Republic of China\\
% ...
$^{4}$ Bochum Ruhr-University, D-44780 Bochum, Germany\\
% ...
$^{10}$ G.I. Budker Institute of Nuclear Physics SB RAS (BINP), Novosibirsk 630090, Russia\\
% ...
$^{15}$ Helmholtz Institute Mainz, Johann-Joachim-Becher-Weg 45, D-55099 Mainz, Germany\\
% ...
$^{42}$ Southeast University, Nanjing 211100, People's Republic of China\\
$^{43}$ State Key Laboratory of Particle Detection and Electronics, Beijing 100049, Hefei 230026, People's Republic of China\\
% ...
$^{55}$ University of Science and Technology of China, Hefei 230026, People's Republic of China\\
% ...
$^{58}$ (A)University of Turin, I-10125, Turin, Italy; (B)University of Eastern Piedmont, I-15121, Alessandria, Italy; (C)INFN, I-10125, Turin, Italy\\
$^{59}$ Uppsala University, Box 516, SE-75120 Uppsala, Sweden\\
% ...
$^{d}$ Also at the Novosibirsk State University, Novosibirsk, 630090, Russia\\
% ...
}\end{center}

\vspace{0.4cm}
\end{small}
    \end{titlepage}	

		\begin{abstract}
			Using data collected with the BESIII detector operating at the Beijing Electron Positron Collider, we search for the process $e^+e^-\rightarrow \eta_c\eta\pi^+\pi^-$. The search is performed using five large data sets recorded at center-of-mass energies of 4.23, 4.26, 4.36, 4.42, and 4.60~GeV. The $\eta_c$ meson is reconstructed in 16 exclusive decay modes. No signal is observed in the $\eta_c$ mass region at any center-of-mass energy. The upper limits on the reaction cross sections are determined to be \uplimone, \uplimtwo, \uplimthree, \uplimfour and \uplimfive pb at the 90\% confidence level at the center-of-mass energies listed above. 
		\end{abstract}
	
	\maketitle
	
	The BESIII collaboration has reported a charmonium like state $Z_c(3900)^{\pm}$ decaying into $J/\psi\pi^\pm$ in the reaction $e^+e^-\rightarrow J/\psi \pi^+\pi^-$
	at $\sqrt{s}$ = 4.26~GeV \cite{Zc3900BES}. This resonance
	 was also observed by the Belle experiment \cite{Belle:Z3900} and confirmed using CLEO-c data \cite{Xiao:Zc3900}. In the CLEO-c dataset, evidence was found for the neutral state $Z_c(3900)^{0}$ in $e^+e^-\rightarrow J/\psi \pi^0\pi^0$~\cite{Xiao:Zc3900}, which was later also observed by BESIII~\cite{Ablikim:2015tbp}. The $Z_c(3900)$ can not be explained as a conventional meson, because of its decay to charmonia  and the existence of its charged state.

	An enhancement in the $D\dbs$ system at a mass of 3890.3~MeV/$c^2$ has been observed in the reaction channel $e^+e^-\rightarrow \pi^+ (D\dbs)^-$ at  a center-of-mass energy of 4.26~GeV, which may be identified as the $Z_c(3900)^{\pm}$ \cite{Ablikim:2013xfr}. Because of the observation of this second decay, interpretations favor the resonance to be either a tetra-quark state or a $D$-meson molecule \cite{Zc_mol}. In addition, a further charged resonance $Z_c(4020)^{\pm}$ was found in the subsystem $h_c\pi^{\pm}$ of the reaction $e^+e^-\rightarrow h_c \pi^+\pi^-$, also at  a center-of-mass energy of 4.26~GeV, closely followed by the discovery of its isospin partner the $Z_c(4020)^0$~\cite{PhysRevLett.111.242001,Ablikim:2014dxl}. Furthermore, structures whose poles are compatible with the $Z_c(4020)$ have been observed by the BESIII collaboration in the reactions $e^+e^-\rightarrow \pi^+(D^{*}\dbs)^{-}$ and $e^+e^-\rightarrow \pi^0(D^{*}\dbs)^{0}$ \cite{PhysRevLett.112.132001,PhysRevLett.115.182002}. 

The observations of the isospin triplets  $Z_c(3900)$ decaying to $J/\psi\pi$ and $Z_c(4020)$ decaying to  $h_c\pi$  suggest the possibility of an unobserved triplet of $Z_c^{\pm,0}$ states decaying to $\eta_c\pi^{\pm,0}$ and  an isospin-singlet state decaying to $\eta_c\eta$. Ref.~\cite{Drenska:2010kg} predicts a tetra-quark state in the mass region of 3.7 to 3.9~GeV with $J^{PC}=0^{++}$ that would satisfy this latter hypothesis. The observation of this resonance would, therefore, add important information to the puzzle of new states and would improve the understanding of their internal structure. 

We search for the reaction $e^+e^-\rightarrow \eta_c\eta \pi^+\pi^-$, as any events observed for this process will allow for studies of possible resonant structure in the $\eta_c\eta$  subsystem. The $\pi\pi$ system must have a relative angular momentum of L=1 to conserve C-parity. It is expected that this pion decay proceeds mainly via the $\rho$ resonance (vector dominance model). This leads to suppression of the decay channel due to isospin conservation and, in addition, a limited phase space below center-of-mass energies of 4.3~GeV. 
Any observed events will also allow for studies of possible resonant substructures in  the $\eta_c\pi^{\pm}$ subsystem.  

 \vspace*{-0.5cm}
	
	\subsection{Detector and Monte Carlo Simulation}\label{sec:detector_and_simulation}
	
	The BESIII detector \cite{BESIIIDetector} is located at the BEPCII double-ring $e^+e^-$ collider. The detector consists of a helium-based multi-layer drift chamber (MDC) with a momentum resolution of 0.5\% for charged particles with a transverse momentum of 1~GeV/$c$, a plastic scintillator based time-of-flight (TOF) system with a time resolution of 68~ps in the barrel and 110~ps in the end caps, a CsI(Tl) electromagnetic calorimeter (EMC) with energy resolutions of 2.5\% and 5.0\% for 1~GeV photons in the barrel and end caps respectively, and a multilayer resistive-plate chamber muon-detection system. The BESIII detector operates in a 1 T magnetic field provided by a superconducting solenoid and has a geometrical acceptance of 93\%.
	
	To optimize selection criteria, estimate detector resolution and reconstruction efficiency, Monte Carlo (MC) simulations are used. The simulation of the BESIII detector is based on \textsc{GEANT4} \cite{Geant4} which models the interaction of particles with the detector material. The initial interaction of the $e^+e^-$ system is modeled with \textsc{KKMC} \cite{KKMC} generator which also handles initial state radiation. Subsequent particle decays are generated with \textsc{EVTGEN} \cite{BESEVTGEN}. The generation of final state radiation is handled by PHOTOS \cite{PHOTOS}. In the simulations the signal reaction channel \mydecay is generated according to a phase-space distribution. The $\eta_c$ is reconstructed in the following 16 decay modes: $\pi^+\pi^-K^+K^-$, $2(K^+K^-)$, $2(\pi^+\pi^-)$, $3(\pi^+\pi^-)$, \Ks$K^{\pm}\pi^{\mp}$, \Ks$K^{\pm}\pi^{\mp}\pi^+\pi^-$, $K^+K^-\pi^0$, $K^+K^-\eta$, $\pi^+\pi^-\eta$, $\pi^+\pi^-\pi^0\pi^0$, $2(\pi^+\pi^-)\eta$,  $2(\pi^+\pi^-\pi^0)$, $K^+K^-2(\pi^+\pi^-)$, $p\bar{p}$, $p\bar{p}\pi^0$ and $p\bar{p}\pi^+\pi^-$. The branching ratios for these decays are taken from Ref.~\cite{EtacBR}. The branching ratios of the decays $\pi^0 \to \gamma\gamma$, $\eta \to \gamma\gamma$,  and  \Ks\ $ \to \pi^+\pi^-$ are taken from the Particle Data Group (PDG) \cite{PDG2016}. For the optimization of the suppression of background reactions various simulated data sets are used, e.g. samples containing light quark and open charm and charmonium final states as well as $e^+e^-$ or $\mu^+\mu^-$ MC samples.
	
	\subsection{Data Analysis}\label{sec:data_analysis}
	
	The search for the reaction is performed at five different center-of-mass energies. The integrated luminosity of these data sets is given in Table~\ref{tab:lumis}.
	\begin{table*}[htb]
		\centering
		\caption{Integrated luminosity of the used data samples and sum over all $\eta_c$ final states of the products of the efficiencies and branching ratios  at the different center-of-mass energies. The center-of-mass energies are taken from Ref. \cite{lumonisities}.}
		\label{tab:lumis}
		\begin{tabular}{c >{\centering\arraybackslash}p{4.cm} >{\centering\arraybackslash}p{4.cm}}
			\hline
			\hline
			$\sqrt{s}$ [MeV] & Luminosity [pb$^{-1}$] & $\sum_{X=1}^{16} \varepsilon_\mathit{tot, X} \mathcal{B}(\eta_c \rightarrow X)$ [\%]\\
			\hline
			4225.54 $\pm$ 0.65 & 1091.7 & 2.07\\
			4257.43 $\pm$ 0.66 & 825.7 & 2.10\\
			4358.26 $\pm$ 0.62 & 539.8 & 2.23\\
			4415.58 $\pm$ 0.64 & 1073.6 & 2.27\\
			4599.53 $\pm$ 0.76 & 566.9 & 2.39\\	
			\hline
			\hline
		\end{tabular}
	\end{table*}
	During event reconstruction, the charged tracks are required to have a point of closest approach to the interaction point within a cylinder with a radius of $V_{xy} =$ 1~cm in the $x$-$y$ plane and a length of $V_z = \pm$10~cm along the beam axis. In addition, the polar angle with respect to the beam axis has to be in the acceptance of the MDC, corresponding to $|\text{cos}(\theta)| < 0.93$. For tracks originating from the decay of long lived particles like the \Ks\, meson, the $V_{xy}$ requirement is omitted while $V_z$ is increased to $\pm$20~cm.
	For each event, the net charge of all reconstructed tracks has to be zero. For particle identification, the joint probability from the energy loss in the MDC and the time-of-flight information of the TOF system is calculated for each particle species ($\pi$, $K$, $p$) and compared for the selection.  Photons are reconstructed from clusters in the EMC. To suppress noise in the EMC, the reconstructed photon energy has to be larger than 25~MeV for $|\text{cos}(\theta)| < 0.80$ and 50~MeV for $0.84 < |\text{cos}(\theta)| < 0.92$. Furthermore, the time difference between the event-start time and the EMC timing information has to be 0~ns $\leq t_{\text{EMC}} \leq$ 700~ns. To suppress clusters formed by split-off photons from charged particle tracks, the angle between a cluster in the EMC and the closest charged track has to be at least \ang{10}. The number of reconstructed photons has to be at least equal to the number of photons expected for the final state in question.
	
	$\pi^0$ and $\eta$ mesons are reconstructed from combinations of photon pairs. To select $\pi^0$ candidates, the invariant mass of two photons must satisfy 
	110~MeV/$c^2 \leq m_{\gamma\gamma} \leq$ 150~MeV/$c^2$ while for $\eta$ candidates the invariant mass has to be in range 500~MeV/$c^2 \leq m_{\gamma\gamma} \leq$ 570~MeV/$c^2$. Candidate \Ks\, mesons are reconstructed by applying a vertex fit to all pairs of oppositely charged particles assuming a pion hypothesis, but requiring no particle identification criteria. For these pairs the decay length $L$ and its uncertainty $\sigma_L$ are calculated from the decay vertex and the primary vertex position. The pair is kept as a \Ks\ candidate if the $\chi^2_{K^0_S}$ of the fit is smaller than 100. In addition, the decay length of the \Ks\, candidate has to satisfy $L/\sigma_L > 2$. Finally, the invariant mass of the pion pair must lie within 15~MeV/$c^2$ of the nominal \Ks\ mass.

	A vertex fit is applied to events passing these criteria, excluding the tracks originating from \Ks\, candidates. In the cases that the vertex fit converges, a kinematic fit is performed to improve the momentum resolution. The fit is constrained by the initial four momentum of the $e^+e^-$ pair and a mass constraint on the $\eta$ mass. If the final state of the $\eta_c$ contains additional $\pi^0$, $\eta$ or \Ks\, mesons, mass constraints are applied on the invariant masses of their daughter particles as well. The selection criteria on the $\chi^2$ value from the kinematic fit is used to suppress poorly reconstructed events and is chosen for each final state to retain 90\% of the signal events. The kinematic fit is not able to discriminate between pions from the initial reaction and pions from the subsequent $\eta_c$ decay as the total four momentum is identical for these two hypotheses. This can lead to multiple candidates per event for the whole reconstruction, with each candidate having the same $\chi^2$. In these cases all candidates are kept for further analysis. It was checked with signal MC data sets that these candidates which have the wrong assignment do not contribute to the signal yield as the form a smooth background distribution. Also all other background distributions show a smooth behaviour at the signal region.
	
	\subsection{Cross Section and Upper-Limit Determination}
	
\begin{figure*}[htb]
		\centering
		\includegraphics[scale=0.8]{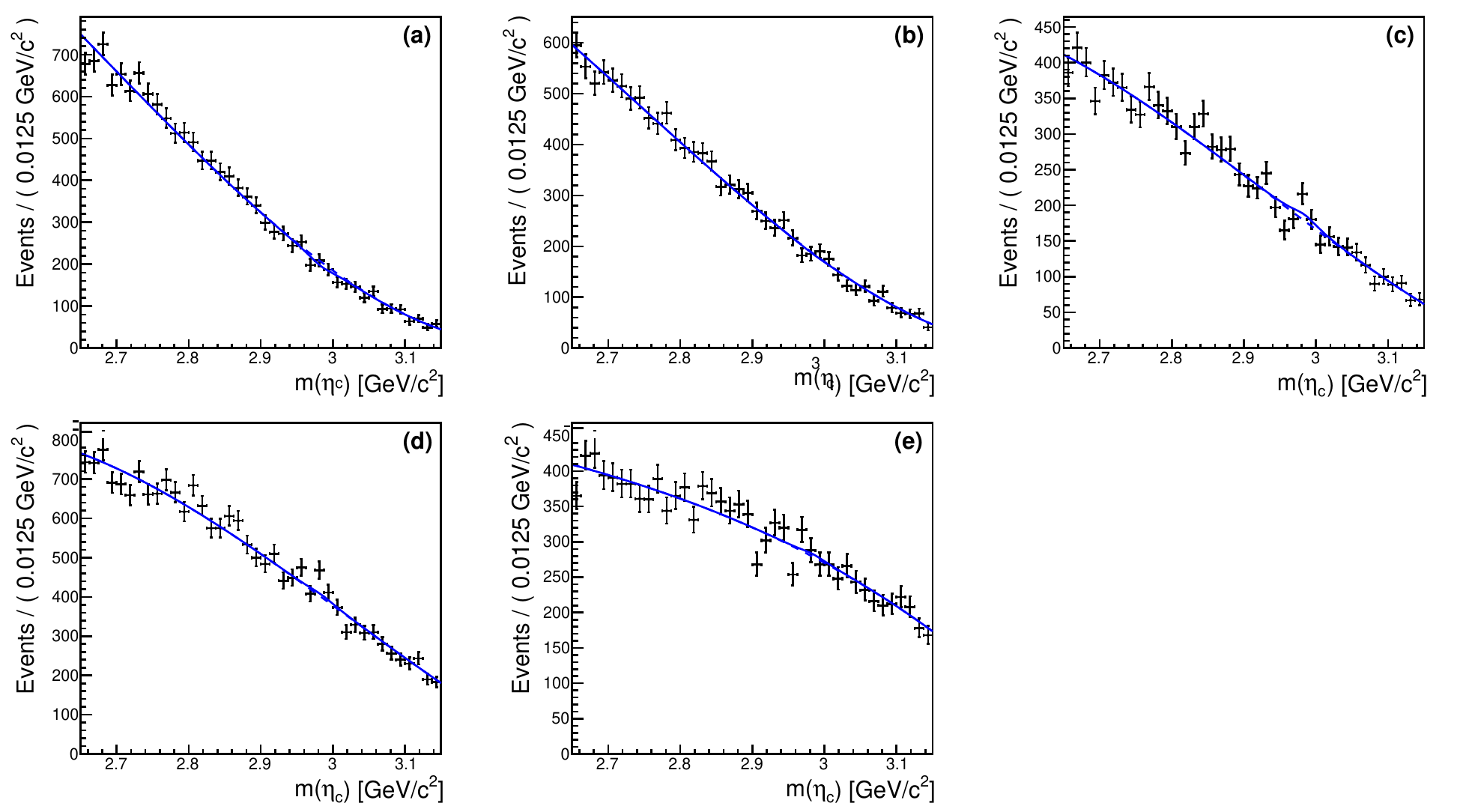}
		\caption{Result of the simultaneous fit to 16 $\eta_c$ decay modes. Shown is the sum of all modes at $\sqrt{s}$ of 4.23~GeV (a), 4.26~GeV (b), 4.36~GeV (c), 4.42~GeV (d), and 4.60~GeV (e). Black points are data, blue line is the sum of the fitting functions.}
		\label{fig:simfit}
	\end{figure*}

	To determine the total event yield of the reaction channel \mydecay a simultaneous extended un-binned maximum-likelihood fit is performed on the invariant-mass distribution $m$ of the $\eta_c$ candidates for all $\eta_c$ decay modes. The fit function $f_{\text{fit}}$ is given by 
	\begin{linenomath}
	\begin{eqnarray*}
		f_{\text{fit}}(m\,|\,n_\text{s}, n_\text{b}, a_n) = &\\  \\
		n_{\text{s}} B(m\,|\,\mu^{\text{PDG}}_{\eta_c}, \Gamma&^{\text{PDG}}_{\eta_c})\otimes G(m\,|\,\mu, \sigma)\\
		 +\,\,\,n_{\text{b}} \sum_{k=1}^{n} a_kT_k(m)
	\end{eqnarray*}
	\end{linenomath}
	for each $\eta_c$ decay mode separately. The signal shape for each decay mode is described by a Breit-Wigner function $B$, whose parameters are fixed to the nominal mass $\mu^{\text{PDG}}_{\eta_c}$ and width $\Gamma^{\text{PDG}}_{\eta_c}$ of the $\eta_c$ meson from the PDG~\cite{PDG2016}. This function is convolved with a Gaussian function $G$ with mean $\mu$ and standard deviation $\sigma$ to account for the detector resolution, which is extracted from signal MC simulation. The number of background events (combinatorial and physical), $n_b$, for each $\eta_c$ decay mode is determined simultaneously in the fit.
For the majority of $\eta_c$ decay modes
the background is described by a $n^{th}$-order  Chebychev polynomial function where the single terms $T_k$ are weighted by the coefficients $a_k$.  For certain decay modes ($\eta_c \rightarrow p\bar{p}$, $K^+K^-\eta$, $K^+K^-\pi^0$ and $K_S^0K^{\pm}\pi^{\mp}$) it is found that an exponential background function provides a better description of the background distribution. The number of signal events, $n_s$, in each $\eta_c$ decay mode is related to the cross section $\sigma$ via the relation
	\begin{linenomath}
	\begin{equation*}
		n_{\text{s}} = \varepsilon_\mathit{tot, X} \mathcal{B}(\eta_c \rightarrow X)\mathcal{B}(\eta \rightarrow \gamma\gamma)\mathcal{L}\sigma
	\end{equation*}  
	\end{linenomath}    
	where $\mathcal{L}$ is the integrated luminosity, $\mathcal{B}(\eta_c \rightarrow X)$ the branching ratio of $\eta_c$ decaying to $X$, and $\varepsilon_\mathit{tot, X}$ the corresponding reconstruction efficiency, which is obtained by fitting the reconstructed $\eta_c$ invariant-mass distribution from signal MC simulation. Table \ref{tab:lumis} shows the sum over all $\eta_c$ final states of the products of the efficiency and branching ratios. 
The invariant mass distribution summed over all decay modes is shown in Fig.\,\ref{fig:simfit}  together with the sum of the fitting curves.
The resulting values for the observed cross section can be found in Table~\ref{tab:result}.
	\begin{table*}[htb]
		\centering
		\caption{Observed cross section $\sigma$ and upper limits (UL) for the reaction \mydecay at the five center-of-mass energies.  UL after all corrections includes the systematic uncertainties plus ISR and vacuum polarization correction.}
		\label{tab:result}
		\begin{tabular}{c  >{\centering\arraybackslash}p{3.cm} >{\centering\arraybackslash}p{2.cm} >{\centering\arraybackslash}p{3.cm} >{\centering\arraybackslash}p{2.5cm}}
			\hline \hline
			\multirow{2}{*}{$E_{c.m.}$ [GeV]} & \multirow{2}{*}{$\sigma$ [pb]} & \multirow{2}{*}{UL [pb]} & UL with systematic uncertainties [pb] & UL after all correction [pb] \\
			\hline 
			4.23 & $-5.39^{+3.15}_{-2.83}$ & 3.5 & 4.2 & 6.2\\
			4.26 & $-0.98^{+4.11}_{-3.53}$ & 6.8 & 7.3 & 10.8\\
			4.36 & $\phantom{-}8.59^{+6.72}_{-6.03}$ & 17.9 & 18.5 & 27.6\\
			4.42 & $\phantom{-}3.07^{+5.36}_{-5.12}$ & 11.2 & 15.2 & 22.6\\
			4.60 & $\phantom{-}3.16^{+6.91}_{-6.51}$ & 14.1 & 15.9 & 23.7\\
			\hline
			\hline
		\end{tabular}
	\end{table*}
	The uncertainties are purely statistical and obtained by a likelihood scan using the MINOS tool \cite{Minos}.
	
\begin{figure*}[htb!]
		\centering
		\includegraphics[scale=0.8]{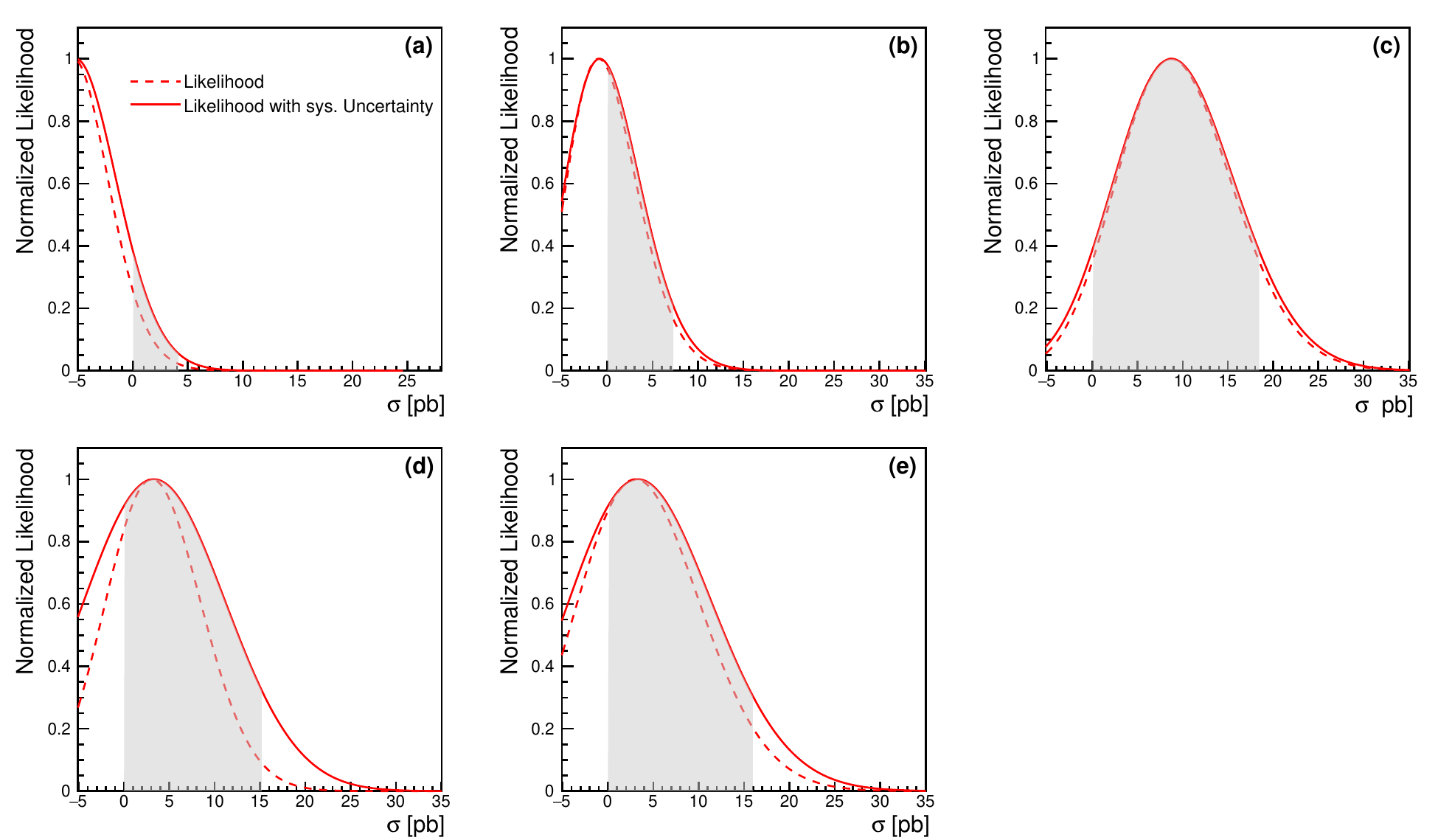}
		\caption{Likelihood curves convoluted with the Gaussian function representing the systematic uncertainties as a function of the cross section at $\sqrt{s}$ of 4.23~GeV (a), 4.26~GeV (b), 4.36~GeV (c), 4.42~GeV (d), and 4.60~GeV (e). The interval correponding to the upper limit at 90\% confidence level is indicated as gray area.}
		\label{fig:upperLimitsSys}
	\end{figure*}

	As no significant signal is observed, an upper limit on the cross section is calculated. For this calculation a Bayesian approach is used. For the prior distribution $\pi(\sigma)$, we
	assume that it is zero for negative values of the cross section and follows a flat distribution otherwise. With this assumption the upper limit is given by
	\begin{linenomath}
	\begin{equation*}
		C(\sigma^{up}) = \frac{\int\limits_{-\infty}^{\sigma^{up}} L(\sigma) \pi(\sigma) d\sigma}{\int\limits_{-\infty}^{\infty} L(\sigma) \pi(\sigma) d\sigma} = \frac{\int\limits_{0}^{\sigma^{up}} L(\sigma) d\sigma}{\int\limits_{0}^{\infty} L(\sigma) d\sigma}.
	\end{equation*}
	\end{linenomath}
	where $L$ is the likelihood function of the simultaneous fit as depicted in Fig.~\ref{fig:upperLimitsSys}. The derived upper limits at 90\% confidence level can also be found in Table~\ref{tab:result}.

	\subsection{Systematic Uncertainties}\label{sec:systematics}
	
	\begin{table*}[htb]
		\centering
		\caption{Total systematic uncertainty at the studied center-of-mass energies.}
		\label{tab:totalUncertainty}
		\begin{tabular}{c >{\centering\arraybackslash}p{1.7cm} >{\centering\arraybackslash}p{1.7cm}>{\centering\arraybackslash}p{1.7cm}>{\centering\arraybackslash}p{1.7cm}>{\centering\arraybackslash}p{1.7cm}}
			\hline
			\hline
			source & $\sigma^{\GeV{4.23}}_\mathrm{sys}$  & $\sigma^{\GeV{4.26}}_\mathrm{sys}$ & $\sigma^{\GeV{4.36}}_\mathrm{sys}$ & $\sigma^{\GeV{4.42}}_\mathrm{sys}$ & $\sigma^{\GeV{4.60}}_\mathrm{sys}$\\
			\hline
			Fit Range [pb]& 0.23 & 0.57& 0.79 & 0.51 & 0.05 \\
			Background Shape [pb] & 1.53 & 1.43 &  0.48 & 5.59 & 3.97 \\
			$\eta_c$ parameters [pb] & 0.15 & 0.06 & 0.16 & 0.05 & 0.05 \\
			$\eta_c$ branching ratio [\%] & 21.6 & 62.6 & 18.0 & 50.2 & 39.6 \\
			Reconstruction Efficiency [\%] & 12.8 & 23.5 & 11.3 & 12.0 & 13.8 \\
			Kinematic Fit [pb] & 0.03 & 0.03 & 0.11 & 0.02& 0.03\\
			Luminosity [\%]& 1.0 & 1.0 & 1.0& 1.0& 1.0\\
			\hline
			total [pb] & 2.05 & 1.67 & 2.04 & 5.83 & 4.19\\
			\hline
			\hline
		\end{tabular}
	\end{table*} 
	
	There are several sources of possible systematic bias in the analysis, for which uncertainties are assigned. These originate in discrepancies in the detector description between MC simulation and data, the knowledge  of the $\eta_c$ branching ratios, the knowledge of the resonance parameters of the $\eta_c$, the kinematic fit, and the background model and fit range used in the simultaneous fit. The systematic uncertainties are summarized in Table \ref{tab:totalUncertainty}.
	 
	The uncertainty associated with the understanding of the  track reconstruction in the MDC is studied with the decays $J/\psi\rightarrow p\bar{p}\pi^+\pi^-$ and $J/\psi\rightarrow \rho\pi$ and is found to be 1\% per charged track \cite{TrackSys}. An additional 1\% per track is applied to account for the knowledge of the particle identification performance~\cite{PID2019}. The systematic uncertainty on the photon detection is estimated using the control samples $\psi(3680) \rightarrow \pi^+\pi^-J/\psi$ with $J/\psi \rightarrow \rho^0\pi^0$ and is determined to be smaller than 1\% for each photon \cite{PhotonEfficiency}. The systematic uncertainty on the \Ks\, reconstruction is estimated to be 1.2\% using the control samples $J/\psi\rightarrow K^*(892)^{\pm}K^{\mp}$ with $K^*(892)^{\pm}\rightarrow K_S^0\pi^{\pm}$ and $J/\psi \rightarrow \phi K_S^0 K^{\pm}\pi^{\mp}$ \cite{KsEfficiency}. The systematic uncertainty associated with  $\pi^0$ and $\eta$ reconstruction is estimated to be 1\% per $\pi^0/\eta$, following the studies reported in Ref. \cite{PiEtaEfficiency} using the control samples $J/\psi\rightarrow \pi^+\pi^-\pi^0$ and $J/\psi\rightarrow \eta p\bar{p}$.  The influence of these uncertainties on the cross section extracted from the simultaneous fit is estimated by multiplying the reconstruction efficiency of each $\eta_c$ final state $X$ with a correction factor $\alpha_X$, which is given by
	\begin{linenomath}
	\begin{equation*}
	 \alpha_X = (\kappa_T)^{n_T}\, (\kappa_\gamma)^{n_\gamma}\, (\kappa_{\pi^0/\eta})^{n_{\pi^0/\eta}}\, (\kappa_{K^0_S})^{n_{K^0_S}} .
	\end{equation*}
	\end{linenomath}
Each $\kappa_Y$ (with $Y = T$ for tracks, $\gamma$ for photons, $\pi^0/\eta$ and $K^0_S$  for the reconstructed mesons)  follows a Gaussian distribution centered at one and a width set to the corresponding uncertainty, while $n_Y$ is the number of tracks, photons etc. in each final state $X$. The simultaneous fit is performed 1000 times while changing the values for $\kappa_Y$ for each fit. The width of the resulting distribution normalized to the extracted cross-section is taken as the systematic uncertainty of the reconstruction efficiency.
	
	The $\eta_c$ branching ratios entering the simultaneous fit are derived from the BESIII measurement in Ref.~\cite{EtacBR}, using the following relation
	%
	\begin{equation*}
		\mathcal{B}(\eta_c \rightarrow X) = \frac{\mathcal{B}(\psi(3680) \rightarrow \pi^0 h_c;  h_c \rightarrow \gamma \eta_c; \eta_c \rightarrow X)}{\mathcal{B}(\psi(3680) \rightarrow \pi^0 h_c;  h_c \rightarrow \gamma \eta_c)}.
	\end{equation*}
	%
	Here the branching ratio $\mathcal{B}(\psi(3680) \rightarrow \pi^0 h_c;  h_c \rightarrow \gamma \eta_c)$ is obtained by combining two measurements performed by BESIII \cite{hcBr1} and CLEO \cite{hcBr2}. To estimate the systematic uncertainty of the branching ratios for the $\eta_c$ final states  a random number is drawn  from a Gaussian distribution whose width is set to the total uncertainty of the combined measurement of the common denominator, and one for each of the 16 modes in the numerator separately. The simultaneous fit is performed again with the updated branching ratios. This is repeated 1000 times and the width of the obtained cross-section distribution normalized to the extracted cross-section is taken as systematic uncertainty.
	
	During the simultaneous fit the mean and width of the signal Breit-Wigner distributions are fixed to the values given by the PDG \cite{PDG2016}. To account for the uncertainties on these values, 1000 fits are performed in which new values for the mean and width of the $\eta_c$ are randomly generated from two independent Gaussian probability distributions, with the parameters of these distributions set according to the central values and uncertainties of the PDG.  The standard deviations of the resulting cross-section distributions are assigned as a systematic uncertainty in the measurement.
	
	To improve the agreement of the $\chi^2$ distribution of the kinematic fit  between signal MC and data, the helix parameters of the charged tracks are smeared using the decay $J/\psi \rightarrow \phi f_0(980)$. The systematic uncertainty associated with the kinematic fit is estimated by switching off this correction, repeating the simultaneous fit and assigning the difference in the cross sections as the uncertainty.
	
	The influence of the mass range over which the fit is performed is studied by narrowing and increasing the range of the fit by 5~MeV/$c^2$. The systematic uncertainty is calculated by taking the maximum difference of the nominal fit value and the values obtained by varying the fit range.

	The systematic uncertainty associated with the description of the background shape is estimated by increasing the order of the Chebychev polynomials by one.  For those  cases where the baseline description of the background is an exponential function, this is replaced by a second-order Chebychev polynomial. The difference between the cross section obtained with the new fit and that with the nominal background model is taken as the systematic uncertainty.
	
	The integrated luminosity is determined by using Bhabha events. The systematic uncertainty of the luminosity measurement has been studied in Ref.~\cite{LuminosityUncertainty} and a relative uncertainty of 1\% is assigned for each center-of-mass energy. 

	To include the systematic uncertainties into the calculation of the upper limits, the likelihood is folded with a Gaussian distribution with a width set to the size of the systematic uncertainties
	\begin{linenomath}
		\begin{equation*}
		L_{\text{sys}}(\sigma) = \int\limits_{-\infty}^{\infty} L(\sigma^{\prime}) \cdot G(\sigma^{\prime}|\sigma, \sigma_{\text{sys}}) d\sigma^{\prime}.
		\end{equation*}
	\end{linenomath}
	The likelihood graphs from this procedure are shown in Fig.~\ref{fig:upperLimitsSys}. The upper limits for the observed cross sections including the systematic uncertainty are listed in Table \ref{tab:result}.
	
	To obtain upper limits for the Born cross sections $\sigma_\mathit{Born}$, the observed cross sections have to be corrected for initial-state radiation (ISR) and vacuum polarization. The equation for the number of signal events in each $\eta_c$ decay mode then reads
	\begin{linenomath}
	\begin{equation*}
		n_{\text{s}} = \varepsilon_\mathit{X, tot}
		\mathcal{B}(\eta_c \rightarrow X)\mathcal{B}(\eta \rightarrow \gamma\gamma)\mathcal{L}\delta_{\mathrm{ISR}}\delta_{vp}\sigma_\mathit{Born},
	\end{equation*} 
	\end{linenomath}
where $\varepsilon_\mathit{X, tot}$ is the total efficiency for final-state $X$, $\mathcal{L}$ is the integrated luminosity,  $\delta_\mathrm{ISR}$ and $\delta_{vp}$ are the correction factors for initial-state radiation  and vacuum polarization, respectively.  The ISR correction factor is given by
	\begin{equation*}
		\delta_\mathrm{ISR} = \int \frac{\sigma(x)}{\sigma_0} \frac{\varepsilon(x)}{\varepsilon_0} W(x) dx.
	\end{equation*}
Here $x$ is the fraction of the beam energy carried away by the ISR photon, $\varepsilon(x)$ the corresponding reconstruction efficiency, $\sigma(x)$ is the line shape of a single resonance, which is assumed to have Breit-Wigner shape in the calculations, and $\sigma_0$ and $\varepsilon_0$ are their counterparts in the absence of initial-state radiation. $W(x)$ is the so-called radiator function~\cite{RadiatorFunction}. The value of $\delta_\mathrm{ISR}$ has a strong dependence on the parameters of the Breit-Wigner line shape,  with the correction being largest for narrow resonances.
As no resonances are observed, we make the conservative assumption that any resonance present has  a width of 10~MeV/$c^2$ which is well below the measured parameters of, for example, the Y(4260). For the determination of the upper limit the most conservative approach is taken by assuming this small resonance is located such that the correction factor is largest.   The value of $\delta_\mathrm{ISR}$ is estimated to be 0.64, independent of the collision energy and the $\eta_c$ final state.    This is shown in Table~\ref{tab:vacuum}, together with the values of $\delta_{vp}$, which are energy dependent and calculated with alphaQED \cite{VacuumCorrection}. The upper limits for the Born cross section are given in the right column of Table~\ref{tab:result}.

	\begin{table}[htb]
		\centering
		\begin{tabular}{c >{\centering\arraybackslash}p{1.5cm} >{\centering\arraybackslash}p{1.5cm}}
			\hline
			\hline
			$E_{c.m.}$ [GeV] & $\delta_{vp}$ & $\delta_\mathrm{ISR}$\\
			\hline
			4.23 & 1.056 & 0.64\\
			4.26 & 1.054 & 0.64\\
			4.36 & 1.051 & 0.64\\
			4.42 & 1.053 & 0.64\\
			4.60 & 1.055 & 0.64\\
			\hline
			\hline
		\end{tabular}
		\caption{Values for the vacuum polarization and ISR corrections for the different data sets. Calculations of the vacuum polarization correction are based on alphaQED \cite{VacuumCorrection}.}
		\label{tab:vacuum}
	\end{table}

	\subsection{Summary and Results}\label{sec:results}

	We perform a search for the process \mydecay at $\sqrt{s} =$ 4.23, 4.26, 4.36, 4.42, and 4.60~GeV with data collected by the BESIII detector. The cross section at each center-of-mass energy is determined by a simultaneous fit to the invariant mass of the $\eta_c$ meson for 16 decay modes. The observed cross sections are determined to be 
	\begin{linenomath}
	\begin{align*}
		\sigma_{\GeV{4.23}} &= -5.39^{+3.15}_{-2.83}\pm 2.05\, \text{pb}\\
		\sigma_{\GeV{4.26}} &= -0.98^{+4.11}_{-3.53}\pm 1.67\,\text{pb}\\
		\sigma_{\GeV{4.36}} &= \phantom{-}8.59^{+6.72}_{-6.03}\pm 2.04\, \text{pb}\\
		\sigma_{\GeV{4.42}} &= \phantom{-}3.07^{+5.36}_{-5.12}\pm 5.83\, \text{pb}\\
		\sigma_{\GeV{4.60}} &= \phantom{-}3.16^{+6.91}_{-6.51}\pm 4.19\, \text{pb}.
	\end{align*}
	\end{linenomath}
	where the first uncertainty is statistical, and the second systematic.  As no significant signal is observed, upper limits on the Born cross sections are determined to be 
	\begin{linenomath}
	\begin{align*}
		\sigma^{\text{up}}_{\GeV{4.23}} &= 6.2\, \text{pb}\\
		\sigma^{\text{up}}_{\GeV{4.26}} &= 10.8\, \text{pb}\\
		\sigma^{\text{up}}_{\GeV{4.36}} &= 27.6\, \text{pb}\\
		\sigma^{\text{up}}_{\GeV{4.42}} &= 22.6\, \text{pb}\\
		\sigma^{\text{up}}_{\GeV{4.60}} &= 23.7\, \text{pb}
	\end{align*}
	\end{linenomath}
	at the 90\% confidence level. These upper limits are of the same order of magnitude as the measured cross sections of the processes $e^+e^-\rightarrow J/\psi\pi^+\pi^-$ and $e^+e^-\rightarrow h_c\pi^+\pi^-$ \cite{PhysRevLett.111.242001, JPsiPiPiCrossSection}.   As no significant  \mydecay signal is seen in the current data set 
it is not yet possible to conclude about possible resonant structures in the final-state subsystems.

	\subsection*{Acknowledgments}
	
The BESIII collaboration thanks the staff of BEPCII and the IHEP computing center for their strong support. This work is supported in part by National Key Basic Research Program of China under Contract No. 2015CB856700; National Natural Science Foundation of China (NSFC) under Contracts Nos. 11625523, 11635010, 11735014; National Natural Science Foundation of China (NSFC) under Contract No. 11835012; the Chinese Academy of Sciences (CAS) Large-Scale Scientific Facility Program; Joint Large-Scale Scientific Facility Funds of the NSFC and CAS under Contracts Nos. U1532257, U1532258, U1732263, U1832207; CAS Key Research Program of Frontier Sciences under Contracts Nos. QYZDJ-SSW-SLH003, QYZDJ-SSW-SLH040; 100 Talents Program of CAS; INPAC and Shanghai Key Laboratory for Particle Physics and Cosmology; German Research Foundation DFG under Contracts Nos. Collaborative Research Center CRC 1044, FOR 2359; Istituto Nazionale di Fisica Nucleare, Italy; Koninklijke Nederlandse Akademie van Wetenschappen (KNAW) under Contract No. 530-4CDP03; Ministry of Development of Turkey under Contract No. DPT2006K-120470; National Science and Technology fund; The Knut and Alice Wallenberg Foundation (Sweden) under Contract No. 2016.0157; The Royal Society, UK under Contract No. DH160214; The Swedish Research Council; U. S. Department of Energy under Contracts Nos. DE-FG02-05ER41374, DE-SC-0010118, DE-SC-0012069; University of Groningen (RuG) and the Helmholtzzentrum fuer Schwerionenforschung GmbH (GSI), Darmstadt.
	\vfil
	\bibliography{Bibliography}
	
\end{document}